# Telling Breaking News Stories from Wikipedia with Social Multimedia: A Case Study of the 2014 Winter Olympics

Thomas Steiner[*]
Google Germany GmbH
ABC-Str. 19, 20354 Hamburg, Germany
tomac@google.com

## ABSTRACT

With the ability to watch Wikipedia and Wikidata edits in realtime, the online encyclopedia and the knowledge base have become increasingly used targets of research for the detection of breaking news events. In this paper, we present a case study of the *2014 Winter Olympics*, where we tell the story of breaking news events in the context of the Olympics with the help of social multimedia stemming from multiple social network sites. Therefore, we have extended the application *Wikipedia Live Monitor*—a tool for the detection of breaking news events—with the capability of automatically creating media galleries that illustrate events. Athletes winning an Olympic competition, a new country leading the medal table, or simply the Olympics themselves are all events newsworthy enough for people to concurrently edit Wikipedia and Wikidata—around the world in many languages. The Olympics being an event of common interest, an even bigger majority of people share the event in a multitude of languages on global social network sites, which makes the event an ideal subject of study. With this work, we connect the world of Wikipedia and Wikidata with the world of social network sites, in order to convey the spirit of the *2014 Winter Olympics*, to tell the story of victory and defeat, and always following the Olympic motto *Citius, Altius, Fortius*. The proposed system—generalized for all sort of breaking news stories—has been put in production in form of the Twitter bot @mediagalleries, available and archived at https://twitter.com/mediagalleries.

## Categories and Subject Descriptors

H.5.1 [Information Interfaces and Presentation]: Multimedia Information Systems

## Keywords

Storytelling, social network sites, multimedia, Wikipedia

[*]Thomas Steiner's second affiliation is Université de Lyon, CNRS Université Lyon 1, LIRIS, UMR5205, F-69622

# 1. INTRODUCTION

## 1.1 Brief History of Wikipedia and Wikidata

The free online encyclopedia *Wikipedia*[1] [15] was formally launched on January 15, 2001 by J. Wales and L. Sanger, albeit the fundamental wiki technology and the underlying concepts are older. Wikipedia's direct predecessor was Nupedia [15], a similarly free online encyclopedia, however, that was exclusively edited by experts following a strict peer-review process. Wikipedia's initial role was to serve as a collaborative platform for draft articles for Nupedia. What happened in practice was that Wikipedia rapidly overtook Nupedia as there were no peer-reviews, and it is now a globally successful highly active [18] Web encyclopedia available in 287 languages with overall more than 30 million articles.[2]

*Wikidata*[3] [25] is a free knowledge base that can be read and edited by both humans and bots. As Wikipedia is a truly global effort, sharing non-language-dependent facts like population figures centrally in a knowledge base makes a lot of sense to facilitate international article expansion. The knowledge base centralizes access to and management of structured data, such as references between Wikipedias and statistical information that can be used in articles. Controversial facts such as borders in conflict regions can be added with multiple values and sources, so that Wikipedia articles can, dependent on their standpoint, choose preferred values.

## 1.2 Social Network Sites and Multimedia

In [6], boyd and Ellison define the term *social network site* as follows. *"We define social network sites as web-based services that allow individuals to (1) construct a public or semi-public profile within a bounded system, (2) articulate a list of other users with whom they share a connection, and (3) view and traverse their list of connections and those made by others within the system. The nature and nomenclature of these connections may vary from site to site."* Social network sites commonly allow their users to publish, share, and react or comment on social multimedia files like videos or photos. Mobile devices like smartphones or tablets are omnipresent at all sorts of events, enabling broad multimedia coverage.

## 1.3 Hypotheses and Research Questions

In this paper, we connect the world of Wikipedia and Wikidata with the world of social networks in order to convey the spirit of the *2014 Winter Olympics*. We automati-cally generate different kinds of media galleries for breaking news events around the Olympics and evaluate the media galleries' relevance and their visual aesthetics. While this paper presents a case study of the *2014 Winter Olympics*, the overall objective is to make the system domain-independent. We are steered by the following hypotheses.

---
[1]Wikipedia: http://www.wikipedia.org/
[2]Wikipedia statistics: http://stats.wikimedia.org/
[3]Wikidata: http://www.wikidata.org/

(ℍ1) Social multimedia is suitable for illustrating breaking news events around the *2014 Winter Olympics*.

(ℍ2) Given different kinds of media galleries for the same breaking news event around the *2014 Winter Olympics*, there is always one predictable preferred kind.

(ℍ3) The key learnings from the domain of the *2014 Winter Olympics* can be generalized to other domains.

These hypotheses lead us to the research questions below.

(ℚ1) What breaking news event features determine the relevancy of the corresponding media gallery?

(ℚ2) What factors determine the choice of the preferred media gallery kind for a breaking news event?

The source code of the application developed in the context of this research as well as all generated multimedia social data are available under the terms of the Apache 2.0 license and can be obtained at the URL https://github.com/tomayac/wikipedia-live-monitor.

The remainder of this paper is structured as follows. Section 2 provides background on the tools *Wikipedia Live Monitor* and *Social Media Illustrator* that we have extended in the context of this work. Section 3 introduces the topic of and the motivation for media gallery aesthetics. Section 4 describes the architecture of the present application. Section 5 contains an evaluation and a discussion of the obtained results. Section 6 gives an overview of related work and finally Section 7 closes the paper with an outlook on future work and conclusions.

## 2. ENABLING TOOLS BACKGROUND

With this research, we build on and extend previous research by Steiner *et al.*, notably the open-source applications *Wikipedia Live Monitor*[4] and *Social Media Illustrator*.[5]

### 2.1 Wikipedia Live Monitor

The open-source application *Wikipedia Live Monitor* was introduced by Steiner *et al.* in [20]. It monitors Wikipedia and Wikidata for concurrent edits that it treats as signals for breaking news events. Whenever a human or bot changes an article of any of the 287 Wikipedias,[6] a change event gets communicated by a chat bot over the Wikimedia IRC server (irc.wikimedia.org),[7] so that parties interested in the data can listen to the changes as they happen. For each language version, there is a specific chat room following the pattern "#" + language + ".wikipedia". Wikipedia articles in different languages are highly interlinked. For example, the English article "en:2014_Winter_Olympics" on the Olympics is interlinked with the Russian article "ru:Зимние_Олимпийские_игры_2014". As *Wikipedia Live Monitor* monitors all language versions of Wikipedia and Wikidata in parallel, it exploits this fact to detect *concurrent edit spikes* of Wikipedia and Wikidata article clusters covering the same semantic concept in multiple languages.

The application thus provides us with titles of articles that we can use as search terms for social multimedia content on social network sites. In the example from above, this could be "Зимние Олимпийские игры 2014"(Russian), "2014 Winter Olympics" (English), "Olympische Winterspiele 2014" (German), or "2014ko Neguko Olinpiar Jokoak" (Basque), *etc.*

### 2.2 Social Media Illustrator

*Social Media Illustrator* is a likewise open-source application by Steiner *et al.*, which was introduced in [17, 19]. It provides a social multimedia search framework that allows for searching for and extraction of multimedia data from the social network sites Google+,[8] Facebook,[9] Twitter,[10] Instagram,[11] YouTube,[12] Flickr,[13] MobyPicture,[14] TwitPic,[15] and Wikimedia Commons.[16] In a first step, it deduplicates exact- and near-duplicate social multimedia data based on an algorithm described in [21]. It then ranks social multimedia data by social signals [17] based on an abstraction layer on top of the social network sites mentioned above and, in a final step, allows for the creation of media galleries following aesthetic principles [22] of the two kinds *Strict Order, Equal Size* and *Loose Order, Varying Size*, defined in [19].

We have ported the crucial parts of the source code of *Social Media Illustrator* from the client-side to the server-side, enabling us now to create media galleries at scale and on demand, based on search terms from a patched version of *Wikipedia Live Monitor*.

## 3. MEDIA GALLERY AESTHETICS

A *media gallery* in the context of our task of telling breaking news stories from Wikipedia with social multimedia is hereby defined as a *best-of* compilation of photos, videos, and microposts[17] retrieved from social networks that are related to a given event.

### 3.1 Media Gallery Kinds

For the generation of media galleries, we apply aesthetic principles as defined by Steiner *et al.* in [19, 22], particularly, we aim for the following three visual aesthetic principles: *(i)* a media gallery is called *balanced* if its shape is rectangular, *(ii)* hole-free if there are no gaps from missing multimedia data, and *(iii)* order-respecting if multimedia data appear in insertion order. Two kinds of media galleries were found [19] to fulfill these principles.

---

[4]Wikipedia Live Monitor: http://wikipedia-irc.herokuapp.com/
[5]Social Media Illustrator: http://social-media-illustrator.herokuapp.com/
[6]List of Wikipedias by size: http://meta.wikimedia.org/wiki/List_of_Wikipedias
[7]Raw IRC feeds of recent changes: http://meta.wikimedia.org/wiki/IRC/Channels#Raw_feeds
[8]Google+: https://plus.google.com/
[9]Facebook: https://www.facebook.com/
[10]Twitter: https://twitter.com/
[11]Instagram: http://instagram.com/
[12]YouTube: http://www.youtube.com/
[13]Flickr: http://www.flickr.com/
[14]MobyPicture: http://www.mobypicture.com/
[15]TwitPic: http://twitpic.com/
[16]Wikimedia Commons: http://commons.wikimedia.org/wiki/Main_Page
[17]A micropost is defined as a textual status message on social network sites, optionally accompanied by multimedia data

*Strict Order, Equal Size.*

A media gallery kind called *Strict Order, Equal Size* that strictly respects the insertion order uses an algorithm that works by resizing all multimedia data in one row to the same height and adjusting the widths in a way that the aspect ratios are maintained. A row is filled until a maximum row height is reached, then a new row (with potentially different height) starts, *etc.* This media gallery kind is strictly *order-respecting*, *hole-free*, and can be *balanced* by adjusting the number of media items in +1 steps.

*Loose Order, Varying Size.*

A media gallery kind called *Loose Order, Varying Size* uses an algorithm that works by cropping all multimedia data to have square aspect ratios. This allows for organizing the items in a way such that one big square always contains two horizontal blocks, each with two pairs of small squares. The media gallery is then formed by iteratively filling big or small squares until a square is full, and then adding the square to the smallest column. This media gallery kind allows any media item to become big, while still being loosely *order-respecting* and *hole-free*. Bringing the media gallery in a *balanced* state can be harder, as depending on the shape both small and big media items may be required.

## 3.2 Why Different Media Gallery Kinds

The main motivation for the *Loose Order, Varying Size* kind is that certain media items can be featured more prominently by making them big, while still loosely respecting the insertion order. Examples of to-be-featured media items can be videos, media items with faces, media items available in High-Density quality, or media items with interesting details [23]. In contrast to *Loose Order, Varying Size*, *Strict Order, Equal Size* does not require cropping, which allows for multimedia data outside of common aspect rations like 1:1 (square), 3:2 (digital SLRs), or 3:4 (iPhone) to be properly fitted in media galleries.

## 4. APPLICATION ARCHITECTURE

The underlying application *Wikipedia Live Monitor* puts Wikipedia and Wikidata article clusters that cover the same semantic concept in a monitoring loop. Article clusters stay in the monitoring loop until their time-to-live has been reached, *i.e.*, until there are no further edits. Some article clusters upon fulfilling the breaking news conditions as defined in [20] will be reported as breaking news events. We have patched the *Wikipedia Live Monitor* source code to have an additional social multimedia search hook in that case. This social multimedia search hook receives as an input the international titles of all articles in the currently breaking article cluster as well as their URLs. It uses them as search terms for a patched instance of the application *Social Media Illustrator* that runs on the same server as the instance of *Wikipedia Live Monitor*. It is configured to always return two media galleries for one request, one of kind *Strict Order, Equal Size* and the other of kind *Loose Order, Varying Size*. Both generated media galleries are saved to disk for archiving purposes following a naming scheme that includes the media gallery kind, the originating search terms, and the UNIX timestamp. A screenshot of the running application can be seen in Figure 1.

On the server-side, we use a library called node-canvas [8] that implements the HTML5 canvas API [5] on the server. This library allowed us to port the original client-side code used in *Social Media Illustrator* with manageable effort to the server. The canvas

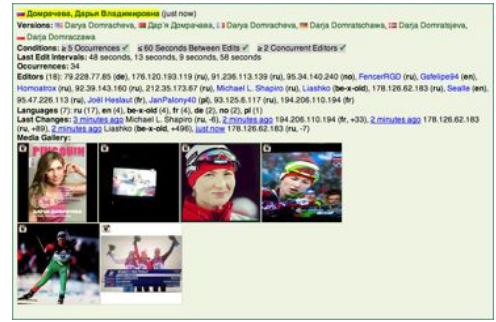

Figure 1: Screenshot of an extended version of Wikipedia Live Monitor running Social Media Illustrator on the server-side with a generated media gallery in HTML format for Darya Domracheva, Women's Boathlon Pursuit event Olympic gold medal winner

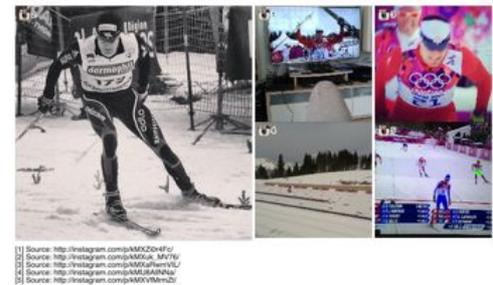

Figure 2: Static dump of a media gallery of kind Loose Order, Varying Size for Dario Cologna, Men's Skiathlon event Olympic gold medal winner

```
<div id="mediaGallery" style="width: 602px;">
  <div class="mediaItem photoBorder" tabindex="1"
      style="width: 307.5px; height: 199.875px;">
    <a href="http://www.flickr.com/[...]">
      <video src="http://www.flickr.com/[...]"
          poster="http://staticflickr.com/[...]"
          class="gallery"
          style="width: 307px; height: 199px;">
      </video>
    </a>
    
  </div>
  [...]
</div>
```

Listing 1: Simplified Strict Order, Equal Size HTML code

API is mainly used in the deduplication algorithm [21] and for creating static dumps of media galleries. Media galleries initially consist of individual images and videos with hyperlinks to the originating microposts as can be seen in Listing 1. For the archived version, a static dump in Portable Network Graphics format with graphical representation of the URLs of the originating microposts is created, an example is shown in Figure 2.

This has the advantage that dumps can be shared as photos over social network sites, in contrast to HTML code with typically expiring URLs that therefore cannot be permanently shared. We share such dumps via the Twitter bot @mediagalleries, available and archived at https://twitter.com/mediagalleries.

# 5. EVALUATION AND DISCUSSION

## 5.1 Quantitative Evaluation

We have evaluated the application based on breaking news events around the *Winter Olympics*[18] that happened between February 8, 20:36 (CET) and February 10, 20:38 (CET), *i.e.*, during an examination period of 48 hours. During this period, 94 unique breaking news events were detected by *Wikipedia Live Monitor*. Uniqueness in this context is defined as events being reported without interruption. For example, *Christof Innerhofer* appeared twice on February 9, once in the morning[19] and once again in the afternoon,[20] so even if the event is about the same person, it is still tracked as two unique breaking news events. Out of these 94 breaking news events, 69 events (≈73%) were related to the *Winter Olympics*. In the 48 hours of our experiment, we have generated overall 804 media galleries, *i.e.*, 402 media galleries of kind *Strict Order, Equal Size* and accordingly 402 kind *Loose Order, Varying Size.* These 402 media gallery pairs covered 67 breaking news events—related to the *Winter Olympics* or not. If we only count the media gallery pairs that were relevant for the *Winter Olympics*, we get 253 media gallery pairs (≈63%) covering 48 of the *Winter Olympics* breaking news events resulting in a **recall of ≈70%** (48 out of 69 of the *Winter Olympics* breaking news events). We calculate the precision in the next subsection.

## 5.2 Qualitative Evaluation

In the qualitative evaluation, we exclusively consider the 253 media gallery pairs that illustrate the 48 *Winter Olympics* breaking news events. We have asked five independent human raters to agree on wether a given media gallery pair is mostly relevant or irrelevant, or completely relevant or irrelevant for the *Winter Olympics* breaking news event in question. By looking at the absolute number of 253 media galleries, 78 media galleries were rated mostly irrelevant or completely irrelevant whereas 175 media galleries were rated mostly relevant or completely relevant. This results in an **absolute precision of ≈69%**. By looking at the number of 48 events related to the *Winter Olympics*, 18 media gallery pairs were rated mostly irrelevant or completely irrelevant whereas 30 media gallery pairs were rated mostly relevant or completely relevant, resulting in a **relative precision of ≈63%**.

## 5.3 Aesthetic Evaluation

In the aesthetic evaluation, we have asked the same five raters for each of the 30 media gallery pairs that they had previously rated mostly relevant or completely relevant to decide on whether the *Strict Order, Equal Size* variant looked better or the *Loose Order, Varying Size* variant. Even after longer discussions between the raters, there was no clear winner, yet each time the raters could tell exactly what bothered them about a given variant. In consequence, rather than providing concrete numbers, we decided to list the raters' most commented-on annoyances. Media galleries of the kind *Strict Order, Equal Size* suffered badly from being unbalanced. Rather than having more social multimedia data, the raters would have rather preferred removing some items in order to get to a balanced state. Media galleries of the kind *Loose Order, Varying Size* suffer less from being unbalanced, but raters consistently remarked positively when they were balanced. The biggest nuisance with *Loose Order, Varying Size* according to our raters were irregular margins that destroy the regular grid pattern, an example thereof can be seen in Figure 3c in the photo of Jamie Anderson with the green-white boarder cap. Overall, raters found *Loose Order, Varying Size* media galleries to be easier to consume, especially when the individual items were diverse.

## 5.4 Temporal Evaluation

In the temporal evaluation, we have examined the effect of evolution of media galleries over time. Figure 3 and Figure 4 show timestamped examples of how the application dynamically recalculates the ranking [17] based on the changing social signals. As social multimedia retrieve interactions on social network sites that we harvest whenever we generate a new media gallery, the algorithm takes these changes into account. Sometimes additional multimedia data appear and previously existing data disappear as is the case in the step from Figure 4d to Figure 4f. Our raters were fascinated by the dynamics of the storytelling where sometimes within well less than a minute media galleries change significantly. They wished for ways to easily navigate back in time in order to relive media gallery evolutions by flipping through the versions.

## 5.5 Discussion

*Reasons for Irrelevant Social Multimedia.*

When we analyzed the kinds of *2014 Winter Olympics* events whose corresponding media galleries our raters had graded as mostly irrelevant or completely irrelevant, we noticed a common pattern. Breaking news events about either the *2014 Winter Olympics* themselves[21] or about any of the Olympic disciplines[22] were illustrated by apparently random Instagram photos. These events consistently have long names with the current year combined with non-ASCII characters. In such cases, the Instagram search API defaults to matching on only parts of the search term rather than the whole phrase, which explains why random photos tagged with the current year, *e.g.*, "#2014 #nofilter" with many social interactions got featured overly prominently. A particularly bad example of a completely irrelevant media gallery can be seen in Figure 5. Following this observation, we have worked around this issue by a better Unicode-aware regular expression.[23]

*Reprise of Hypotheses and Research Questions.*

We have shown that ($\mathbb{H}1$) holds true, social multimedia can successfully illustrate breaking news events around the *2014 Winter Olympics*. In contrast, we had to weaken ($\mathbb{H}2$), as our raters on a case-by-case basis preferred the one or the other kind of media gallery. They always preferred the balanced version of either kind, however, when media galleries were unbalanced, other factors like irregular margins or concrete multimedia contents in general determined their preference and also cropping was not seen as a big issue. There seems to be a slight advantage for more square-like media galleries of either kind, but more profound A/B or multivariate testing is required. Regarding ($\mathbb{H}3$), we have already successfully applied the key learnings concerning aesthetics and search term handling from the *2014 Winter Olympics* domain to other domains and are now able to create relevant and aesthetic media galleries for any kind of breaking news event detected by *Wikipedia Live Monitor*. Looking at our two research questions, for ($\mathbb{Q}1$), there is definitely a tight relation between the category of the breaking news event and the relevancy of the generated media galleries. We have seen in our experiments that events concerning persons throughout reveal highly relevant media galleries. This also holds true for events outside of the *2014 Winter Olympics*. For ($\mathbb{Q}2$), in upcoming versions of our application, we will put additional emphasis on multimedia contents, as we have learned that factors like irregular margins play a key role as they destroy the regular grid structure, which the human eye is very unforgiving of. Our eyes need focal points, visually diverse media galleries with multimedia data of different sizes and contrasting colors can help the eye, albeit our raters also liked color-harmonious media galleries with similar sizes that invite the eye to browse longer.

---

[18] In order to avoid confusion of the numbers, in this section, we strip the 2014 from the event title
[19] Christof Innerhofer (morning): https://twitter.com/WikiLiveMon/status/432438422431887360
[20] Christof Innerhofer (afternoon): https://twitter.com/WikiLiveMon/status/432506953890537472
[21] For example, 2014 Winter Olympics: https://twitter.com/WikiLiveMon/status/432856683036278784
[22] For example, Luge at the 2014 Winter Olympics – Men's singles: https://twitter.com/WikiLiveMon/status/432577721626296320
[23] Instagram customization: http://bit.ly/instagram-unicode

## 6. RELATED WORK

We focus on related work that follows a *holistic approach* for storytelling with social media rather than looking at individual bits and pieces like event detection, media deduplication and clustering, *etc.* Related work can be grouped in several fields.

### *Storytelling with Social Media Around the Olympics.*

The Twitter data journalism team have created a visualization of the most-shared Olympics photos on Twitter [16] that can be filtered by day and country. It is unclear to what extent the system is based on automated hashtag analysis and manual content curation. More visualizations like an interactive map and an athlete follower graph are listed in a post on the Twitter blog [13].

### *Event Archiving and Summarization.*

Event archiving services such as Eventifier[24] do a great job at storing all the social media content around entire events, however, do not currently rank the information. Closest to our approach is Seen[25] an engine that aggregates, organizes, and ranks media and collects information on topics trending in social media. Seen does not support tracking of media gallery evolution. Finally there is MediaFinder [24],[26] that uses a fork of the media item collector used in *Social Media Illustrator* MediaFinder specializes on clustering media items based on named entities.

### *Manual Social Multimedia Curation.*

Examples of manual social media curation tools are FlypSite,[27] a tool that facilitates the creation of embeddable second screen applications or TV social media widgets, Storify [7, 1],[28] a service that lets users create timelines using social media, and Storyful,[29] a news agency focused on verifying and distributing user-generated content from social networks related to news events.

---

[24]Eventifier: http://eventifier.co/
[25]Seen: http://seen.co/
[26]MediaFinder: http://mediafinder.eurecom.fr/
[27]FlypSite: http://www.flyp.tv/
[28]Storify: http://storify.com/
[29]Storyful: http://storyful.com/

### *Identification and Aesthetic Presentation.*

Automated and semi-automated approaches for content identification exist, for example, [10] and [11] by Liu *et al.* who combine semantic inference and visual analysis to automatically find media items that illustrate events. Further, there are [4] and [3] by Becker *et al.* who focus on identifying media items related to events by learning similarity metrics and identifying search terms. However, these event-related social multimedia data identification approaches do not deal with the tasks of ranking [9], deduplicating [26], and representing the event-related content aesthetically [14, 12]. Obrador *et al.* present a photo collection summarization system that includes storytelling principles and face and image aesthetic ranking, however, that is not interactive.

## 7. FUTURE WORK AND CONCLUSIONS

In this paper we have connected the world of breaking news events based on detected concurrent Wikipedia and Wikidata edits with the world of social network sites at the example of the *2014 Winter Olympics*. We have extended the existing tools *Wikipedia Live Monitor* and *Social Media Illustrator* so that now we have an automated means of generating media galleries for breaking news events at scale in the form of the Twitter bot @mediagalleries, available and archived at https://twitter.com/mediagalleries. We have evaluated media galleries that were auto-generated in the context of the Olympics quantitatively, qualitatively, aesthetically, and temporally. From the 69 *Winter Olympics* breaking news events, we could illustrate 48 events, resulting in a recall of successfully illustrated Olympics breaking news events of ≈70%. From these 48 events, 30 events were illustrated with mostly relevant or completely relevant content, resulting in a relative precision of ≈63%, still without the improvements discussed in Subsection 5.5 that are now in place.

Future work will mainly focus on further improving the visual aesthetics of the generated media galleries. While certain aspects like balancedness and overall shape are straight-forward targets to tackle, more subtle issues like analyzing social multimedia data contents for unwanted features (*e.g.*, irregular margins, cropped textual overlays, *etc.*) or wanted features (*e.g.*, faces) that negatively or positively impact the media gallery harmony will require advanced heuristics, especially given our near-realtime demands at the system. Advanced A/B or multivariate tests that optimize on click-through-rate, consumption time, or other variables, and where several media gallery kinds compete against each other, can help reveal new insights about what makes attractive media galleries. The temporal evolution of breaking news stories is another area of research that our application facilitates to explore. A certainly very drastic example are the *Boston Marathon bombings*.[30] The terror attacks were correctly detected by *Wikipedia Live Monitor* [2], potential use cases are in disaster recovery.

Concluding, we are excited by the broad range of possible future use cases and the upcoming improvements of the application that we have unlocked through this present case study on the *2014 Winter Olympics*. By releasing the source code and the social multimedia data created in the context of this research under the permissive Apache 2.0 license, we invite and hope for others to pick up. It is the taking part that counts: https://github.com/tomayac/wikipedia-live-monitor.

---

[30]Boston Marathon bombings: http://en.wikipedia.org/wiki/Boston_Marathon_bombings

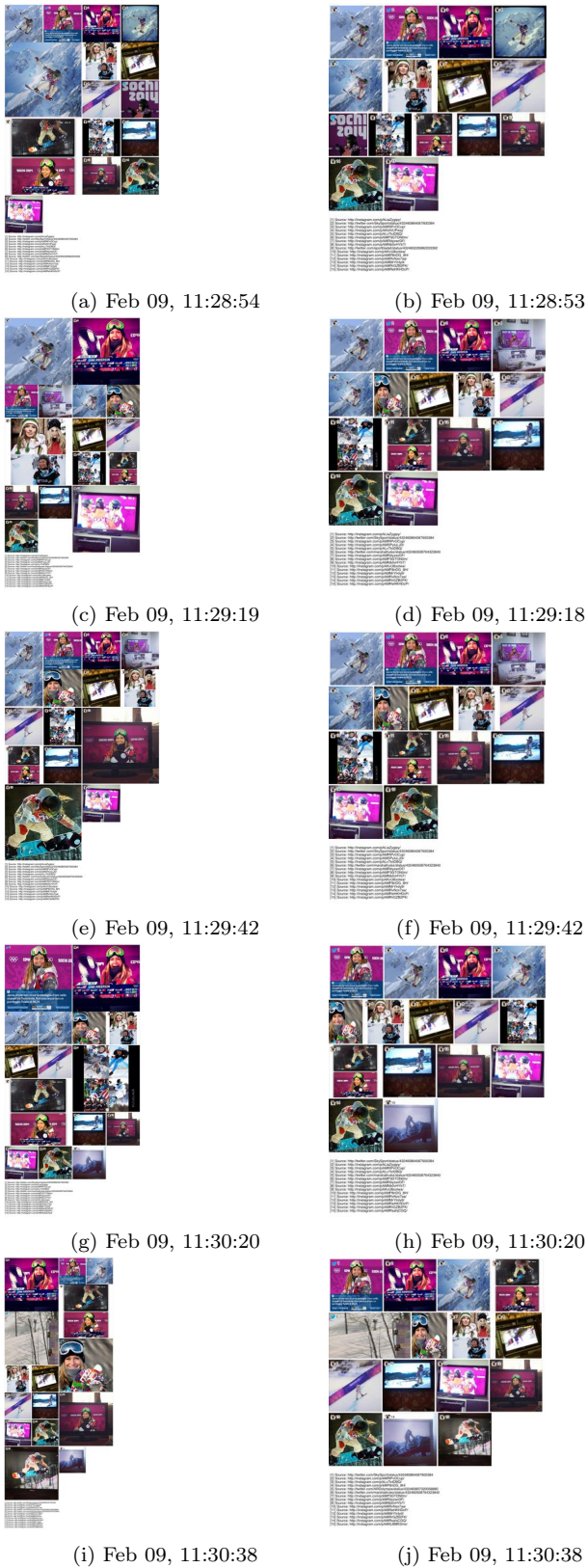

Figure 3: Media galleries for Jamie Anderson, Women's Slopestyle Olympic gold medal winner (times in CET, left: Loose Order, Varying Size, right: Strict Order, Equal Size) https://twitter.com/WikiLiveMon/status/432460983676989440

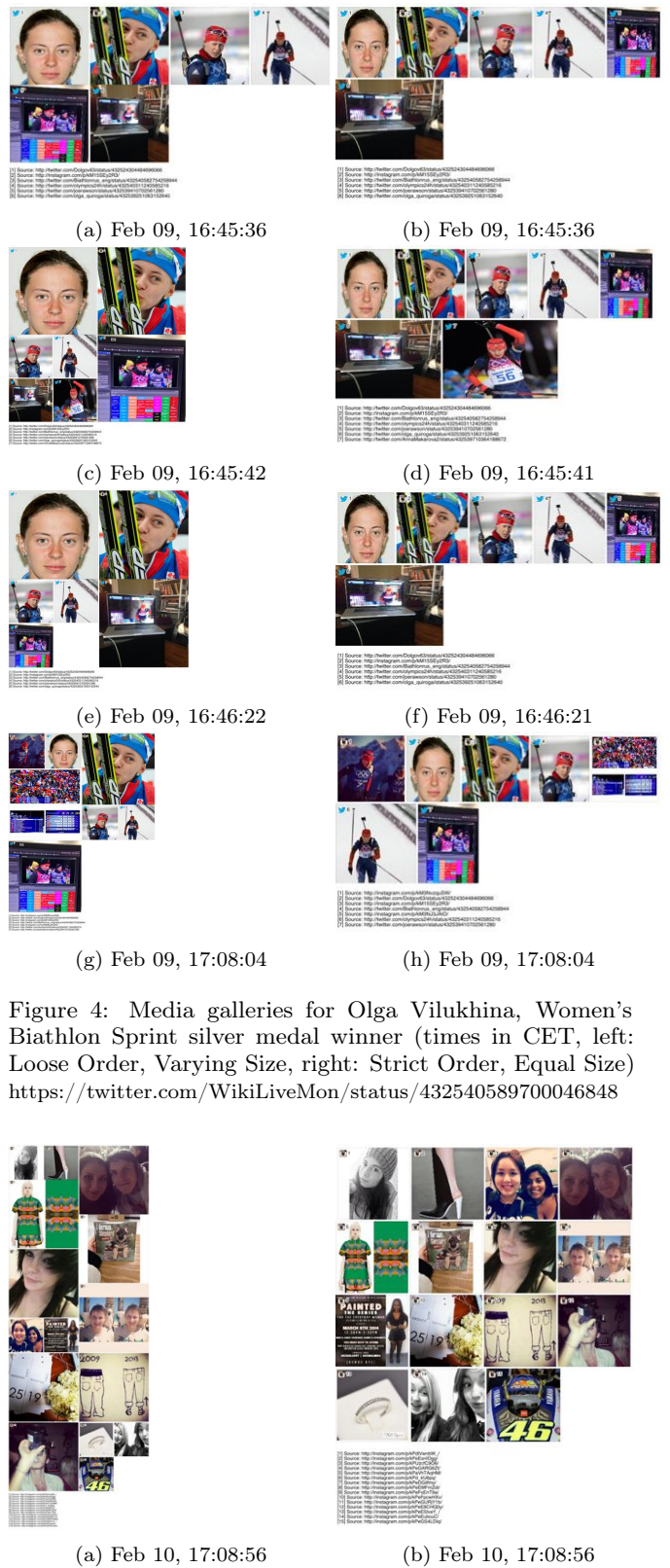

Figure 4: Media galleries for Olga Vilukhina, Women's Biathlon Sprint silver medal winner (times in CET, left: Loose Order, Varying Size, right: Strict Order, Equal Size) https://twitter.com/WikiLiveMon/status/432540589700046848

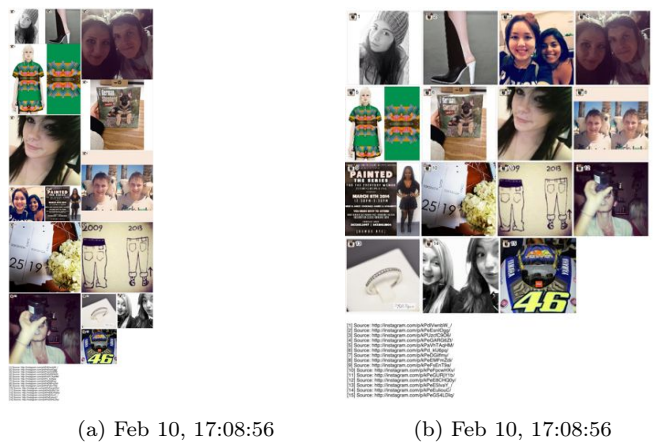

Figure 5: Completely irrelevant media gallery for 2014 Winter Olympics medal table (times in CET, left: Loose Order, Varying Size, right: Strict Order, Equal Size) https://twitter.com/WikiLiveMon/status/432909036313673728